\documentclass{ceab}   

\usepackage{epsfig}     
\usepackage{graphicx}   

\usepackage{ceabbib}     
\usepackage[T1]{fontenc}

\begin{document}

\title{Getting to know the cataclysmic variable beneath the nova eruption}

\author{C. Tappert$^1$, L. Schmidtobreick$^2$, A. Ederoclite$^3$ and 
N. Vogt$^1$
\vspace{2mm}\\
\it $^1$Dpto. de F\'{\i}sica y Astronom\'{\i}a, Universidad de Valpara\'{\i}so,
\\
\it Avda. Gran Breta\~na 1111, Valpara\'{\i}so, Chile\\
\it $^2$European Southern Observatory, Alonso de Cordova 3106, Santiago, Chile\\
\it $^3$Centro de Estudios de F\'{\i}sica del Cosmos de Arag\'on, Plaza San 
Juan 1, \\
\it Planta 2, Teruel, E44001, Spain
}

\maketitle

\begin{abstract}
The eruption of a (classical) nova is widely accepted to be a recurrent
event in the lifetime of a cataclysmic binary star. In-between eruptions the 
system should therefore behave as a ``normal'' cataclysmic variable (CV), 
i.e.~according to its characteristic properties like the mass-transfer rate
or the strength of the magnetic field of the white dwarf. How important are
these characteristics for the nova eruption itself, i.e.~which type of
systems preferably undergo a nova eruption? This question could in principle
be addressed by comparing the post-nova systems with the general CV
population. However, information on post-novae is scarce, even to the extent
that the identification of the post-nova is ambiguous in most cases. In this
paper we inform on the progress of a project that has been undertaken to
significantly improve the number of confirmed post-novae, thus ultimately
providing the means for a better understanding of these objects.
\end{abstract}


\section{Introduction}

A nova eruption in a cataclysmic variable (CV) occurs as a thermonuclear
explosion on the surface of the white dwarf primary star once it has
accumulated a critical mass from its late-type, usually main-sequence,
companion. In the process material is ejected into the interstellar medium,
typically amounting to $10^{-5}$ to $10^{-4}~\mathrm{M_\odot~yr^{-1}}$ 
\citep[e.g.,][]{yaronetal05-1}. Since the system is not destroyed by the
eruption, this is thought to be a recurrent event, with recurrence times
$>10^{3}~\mathrm{yr}$ \citep[see ][]{sharaetal12-3}.

In-between nova eruptions the binary is supposed to appear as a ``normal''
CV, i.e.~its behaviour is dominated by its current mass-transfer rate
and the magnetic field strength of the white dwarf \citep{vogt89-1}.
Furthermore, the hibernation model predicts that most of the time between 
eruptions the system passes as a detached binary 
\citep{sharaetal86-1,prialnik+shara86-1}. While there is still 
no observational evidence for this scenario, i.e.~the state of actual 
``hibernation'' \citep[e.g.,][]{nayloretal92-1}, it is already well established 
that old novae are part of the CV community. For example, the system DQ Her 
(Nova Her 1934) is known as the prototype intermediate polar, while RR Pic 
(Nova Pic 1925) shows the characteristics of an SW Sex CV 
\citep{schmidtobreicketal03-4}. Especially important in this context has been
the discovery of nova shells around the dwarf novae Z Cam and AT Cnc
\citep{sharaetal07-1,sharaetal12-4}, since it proves that (at least some)
original CVs are also old novae. 

All in all, however, our knowledge on old novae is largely incomplete.
There are about 200 reported nova eruptions that occurred before 1980
\citep{downesetal05-1}, but for less than half of them a spectrum of the 
post-nova has been obtained, and about 80 objects even still lack an
unambiguous identification. A first attempt to remedy this situation was 
undertaken by \citet{schmidtobreicketal05-2} who, however, concentrated 
exclusively novae with large eruption amplitudes. Among others, they 
recovered the old nova V840 Oph, an apparently carbon-rich CV 
\citep{schmidtobreicketal03-5} which raises the question if the abundance 
pattern of novae is different from other CVs. 

Furthermore, only for 28 pre-1980 novae the orbital period is well established.
Because the period distribution diagram is one of the most important tools in 
the study of the evolution of compact binaries, this scarcity of respective
information presents a severe obstacle for any systematic study on old novae
and their place within the general CV population. This concerns questions like 
the importance of magnetic fields for the nova eruption (what is the fraction 
of magnetic CVs among novae compared to the general CV population?), the 
mass-transfer rate averaged over long time-scales (is there a bias against 
intrinsically faint systems?), and the hibernation model (do post-novae 
eventually become detached?).

\section{The search for old novae}

\begin{figure}
\includegraphics[width=\columnwidth]{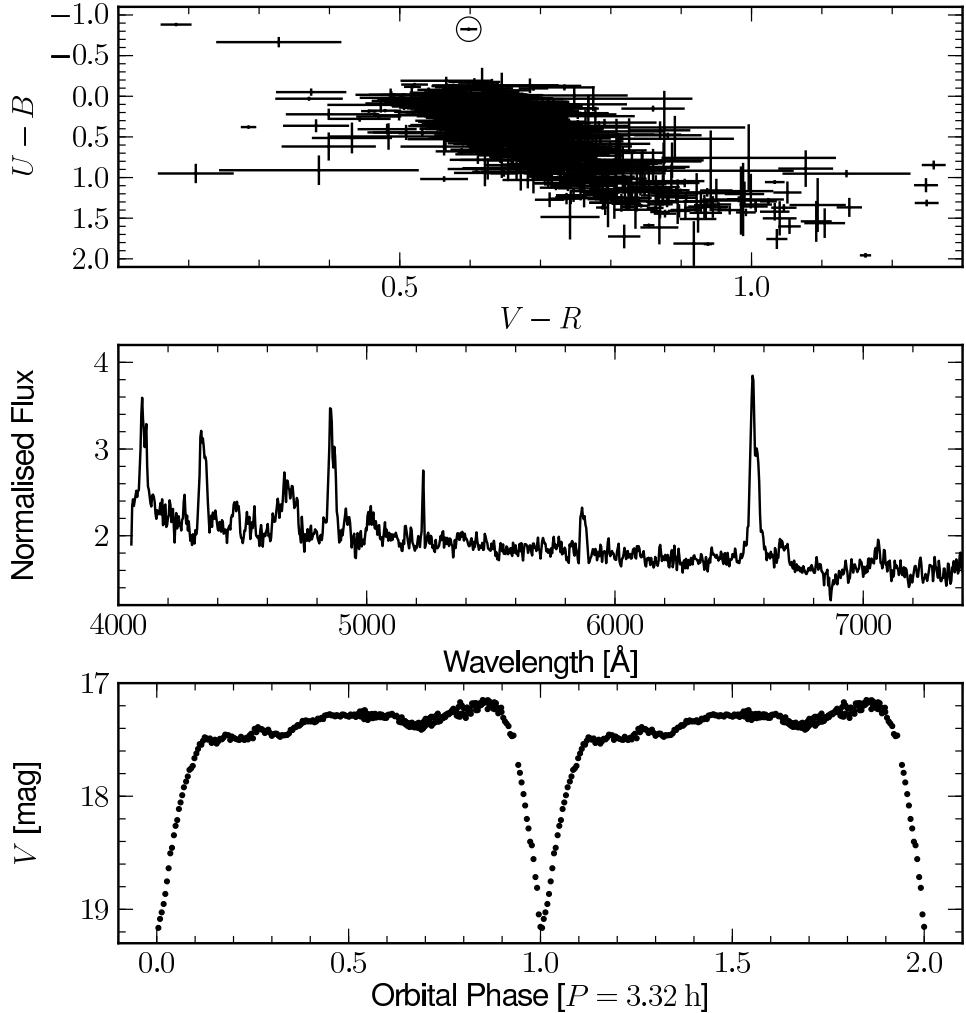}
\caption[]{The data on the old nova V728 Sco, taken in three runs using
EFOSC2 \citep*{eckertetal89-1} on the ESO-NTT, La Silla, Chile. 
{\bf Top:} Colour-colour diagram of the $4.5^\prime \times 4.5^\prime$ field 
centred on coordinates taken from the \citet{downesetal05-1} catalogue. The 
nova is marked by a circle.
{\bf Middle:} Low-resolution spectrum that confirmed the nova candidate.
{\bf Bottom:} Photometric light curve folded on the orbital period of
3.32 h.}
\label{v728data_fig}
\end{figure}

In order to establish a sample of properly identified post-novae we have
begun observations of the nova candidates listed in the \citet{downesetal05-1}
catalogue. We limit our research to novae that were reported to have erupted
at least 30 years ago (i.e.~before 1980) because in most systems the 
contribution of the nova shell in the optical range becomes negligible after
about three decades. First results of this survey have been published in
\citet{tappertetal12-1}.

We use UBVR photometry to select candidates for the post-nova based on their
position in the colour-colour diagram. The typical components of CVs (white
dwarf + red dwarf + accretion disc or stream) place these systems away from 
the main-sequence, the actual position depending strongly on the relation
between the individual contributions. The candidates are then examined with
long-slit spectroscopy for CV characteristic features, like emission lines
of the hydrogen and helium series. Finally for the brightest confirmed
post-novae we attempt to derive the orbital period via time-series spectroscopy
or photometric light curves.

In Fig.~\ref{v728data_fig} we present our study on the old nova V728 Sco as
an example. The system erupted in October 1862 \citep{tebbutt78-3}, putting
it among the five oldest novae in the southern hemisphere. Our UBVR diagram
(top plot) shows the nova well separated from the main-sequence. The long-slit
spectrum presents for a nova unusually strong emission lines of the
Balmer and He{\sc I} series, indicating a comparatively low mass-transfer rate.
The presence of He{\sc II} $\lambda$4686 emission, on the other hand, is 
evidence for a hot component somewhere in the system. The width especially
of the low-excitation lines indicates a high system inclination. The latter
is confirmed by our time-series photometry (bottom plot) that reveals an
eclipsing system with an orbital period of 3.32 h.

\section{The story so far}

\begin{figure}
\includegraphics[width=\columnwidth]{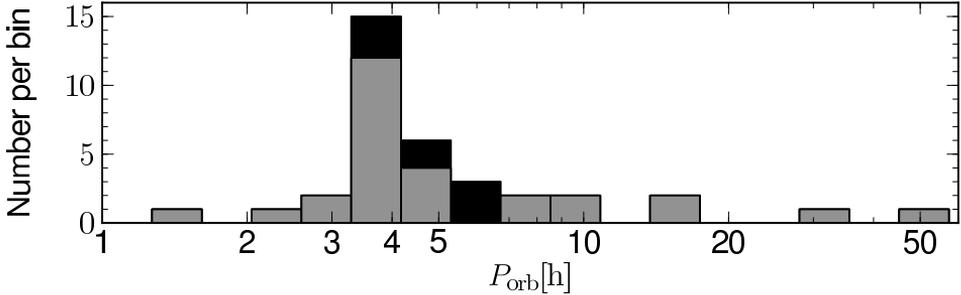}
\caption[]{The period distribution of the pre-1980 novae. New systems are
shown in black.}
\label{novaphis_fig}
\end{figure}

The initial sample of reported pre-1980 nova eruptions in the southern
hemisphere (DEC $<+20$) consisted of 28 novae with known orbital period, 9 
confirmed novae with unknown period, 32 targets without a post-nova spectrum, 
and 86 objects with no proper identification. In the course of our
project we have since then confirmed 13 novae spectroscopically and identified
two variable stars (potentially Miras) whose variability had been mistaken
for a nova eruption \citep[][give details on the majority of these results]%
{tappertetal12-1}. We have furthermore determined the orbital period for
eight novae, which already represents an increase of almost 30\% with 
respect to the previously known periods. Among those eight there are two 
eclipsing novae: V728 Sco, with $P_\mathrm{orb} = 3.32~\mathrm{h}$, and 
V909 Sgr, with $P_\mathrm{orb} = 3.44~\mathrm{h}$. Note that the latter
object had already been reported by \citet{diaz+bruch97-1} to be an eclipsing 
nova with a possible period of 3.36 h.

In Fig.~\ref{novaphis_fig} we show the period distribution of our sample.
The eight additions of our present research further emphasise the apparent
clustering of the orbital periods around 3--6 h, with 2/3 of the novae having 
periods in this range. This is the region where the CVs with the highest 
mass-transfer rates are situated \citep{townsley+gaensicke09-1}. This 
clustering is therefore not unexpected since it appears to support the simple 
idea that the white dwarfs in CVs with high mass-transfer rates accumulate 
the critical mass for a nova eruption faster, leading to shorter eruption 
recurrence times than for low-mass-transfer CVs. However, the current sample 
size of 36 novae is much too small for definite conclusions. 

Our project still very much represents work in progress. New observations
are already underway, with more planned for the future. We therefore expect
to significantly improve on the nova sample in the next years, so that
it can be used for in-depth statistical analyses.

\section*{Acknowledgements} 
This research was supported by FONDECYT Regular grant 1120338 (CT and NV).
The CEFCA is funded by the Fondo de Inversiones de Teruel, supported by both 
the Government of Spain (50\%) and the regional Government of Arag\'on (50\%). 
This work has been partially funded by the spanish Ministerio de Ciencia e 
Innovaci\'on through the PNAYA, under grants AYA2006-14056 and through the 
ICTS 2009-14

We furthermore gratefully acknowledge ample use of the SIMBAD database, 
operated at CDS, Strasbourg, France, and of NASA's Astrophysics Data System 
Bibliographic Services.


\end{document}